\documentclass[aps,showpacs,prb,twocolumn,superscriptaddress,floatfix]{revtex4}
\usepackage{graphicx}

\begin{document}

\title{NMR in the normal and in the superconducting state of MgB$_2$ and comparison with AlB$_2$}

\author{S.~H.~Baek} \affiliation{Department of Physics and Astronomy and Ames
Laboratory, Iowa State University, Ames, IA 50011}
\author{B.~J.~Suh}\affiliation{Department of Physics, the Catholic University of
Korea, Puchon, 420-743, Korea}
\author{E.~Pavarini} \affiliation{Department of Physics ``A. Volta" and Unita'
INFM di Pavia, Via Bassi 6, I27100 Pavia, Italy}
\author{F.~Borsa}  \affiliation{Department of Physics and Astronomy and Ames
Laboratory, Iowa State University, Ames, IA
50011}\affiliation{Department of Physics ``A. Volta" and Unita'
INFM di Pavia, Via Bassi 6, I27100 Pavia, Italy}
\author{R.~G.~Barnes} \affiliation{Department of Physics and Astronomy and Ames
Laboratory, Iowa State University, Ames, IA 50011}
\author{S.~L.~Bud'ko}\affiliation{Department of Physics and Astronomy and Ames
Laboratory, Iowa State University, Ames, IA 50011}
\author{P.~C.~Canfield}\affiliation{Department of Physics and Astronomy and Ames
Laboratory, Iowa State University, Ames, IA 50011}


\date{\today}

\begin{abstract}
$^{11}$B NMR measurements have been performed in $^{11}$B enriched
MgB$_2$ powder samples in external fields of 0.813, 1.55, 4.7 and
7.2 T both in the normal phase and in the superconducting phase. A
previously unreported dipolar Pake doublet has been observed in
the quadrupole perturbed NMR spectrum. The Knight shift can thus
be accurately determined by narrowing the line with the Magic
Angle Spinning (MAS) technique. Results of Knight shift ($K$) and
relaxation rates ($1/T_1$) for both $^{11}$B and $^{27}$Al nuclei
are reported also for AlB$_2$. The comparison of the data in the
two compounds shows the dramatic drop of the density of states at
the boron site in AlB$_2$ with respect to MgB$_2$. The
experimental values for $K$ and $1/T_1$ are in most cases in good
agreement with the theoretical values obtained from first
principles calculations. The recovery of the nuclear magnetization
below $T_c$ in random powder samples is non-exponential due to the
anisotropy of the upper critical field. The exponential drop of
$1/T_1$ in the superconducting phase observed by Kotegawa et
al.~is confirmed here but not the coherence peak.
\end{abstract}


\maketitle

\section{INTRODUCTION}

After the discovery of superconductivity~\cite{naga} in MgB$_2$
with $T_c \sim 39$ K  and subsequent observation of boron isotope
effect~\cite{budko2,hinks} confirming that MgB$_2$ is a
phonon-mediated BCS superconductor, much effort have been devoted
to this intermetallic compound up to date due to its remarkably
high $T_c$ among BCS superconductors.

Nuclear Magnetic Resonance (NMR) is a suitable microscopic tool to
investigate the electronic structure in the normal state, the
structure of the gap and the flux line lattice in the
superconducting state.~\cite{rigamonti} This justifies the large
number of studies by both $^{11}$B and $^{25}$Mg NMR in MgB$_2$
which have already appeared in the
literature.~\cite{jung,tsvya,mali,papa,kotegawa,gera,tou}

Regarding the $^{11}$B NMR, there is substantial agreement about
the measurements of the quadrupole coupling constant $\nu_Q$ and
about the nuclear spin-lattice relaxation rate, $1/T_1$, in the
normal phase. On the other hand  there is considerable controversy
concerning the $^{11}$B Knight shifts in the normal phase and
about the temperature dependence of $1/T_1$ in the superconducting
phase. In the present paper, we report new results for the
$^{11}$B NMR spectrum in the normal phase and an accurate
determination of the Knight shift by using the Magic Angle
Spinning (MAS) technique. We also find an explanation for the
discrepancies in the results of $1/T_1$ below $T_c$ reported by
different authors, based on the strong anisotropy of the upper
critical field in MgB$_2$.

Since NMR in metals probes the density of states (DOS) at the
Fermi level, we have performed measurements of both $^{11}$B and
$^{27}$Al NMR in AlB$_2$ in order to compare the DOS in the two
compounds. Finally the experimental results for Knight shifts and
relaxation rates in both MgB$_2$ and AlB$_2$ have been compared
with \textit{ab-initio} calculations.

\section{Experimental details and sample preparation}

 MgB$_2$ crystallizes in the hexagonal AlB$_2$ type structure,
 which consists of alternating hexagonal layers of Mg atoms and graphite-like honeycomb
 layers of B atoms.
 Powder samples were prepared with the method described in Ref.~[\onlinecite{budko2}].
 X-ray powder diffraction measurements confirmed the hexagonal unit cell of
 MgB$_2$.~\cite{naga,budko2}
 Magnetization measurements done at $H =$ 2.5 mT yield a transition temperature
 $T_c = 39.2$ K with a shielding volume fraction close to 100 \%.~\cite{budko2,finnemore,canfield}
 We have investigated several  samples from different batches of
 polycrystalline $^{11}$B enriched  MgB$_2$ in order to check the reproducibility of the data.
 Also we have performed measurements in samples from the same batch both in bulk and in
 powder ground to different particle size.
 No substantial differences were observed in the NMR measurements in the different samples.
 The onset of superconductivity  was also determined by monitoring the detuning of the NMR
 circuit occurring at the irreversibility temperature, $T_{\mbox{irr}}$.
 This type of measurement corresponds to probing the temperature dependence of the radio
 frequency (rf) surface resistance.~\cite{jung}
 Thus, as the magnetic field is increased the transition region broadens due to the
 dissipation associated with flux line motion below $T_c$, and at 7.2 T
 no detuning can be observed although the magnetization measurements
 indicate a $T_c = 23$ K at 7 T.~\cite{jung}
 The transition temperature at 4.7 T was found to be $T_c = 27.5$ K and at 1.55 T, $T_c =34$ K.

 $^{11}$B and $^{27}$Al NMR and spin-relaxation measurements were performed with home built
 Fourier transform (FT) pulse spectrometers operating at variable frequencies.
 The $\pi/2$ radio frequency (rf) pulse length was typically 6 $\mu$s.
 The Magic Angle Spinning (MAS) experiment was performed with a home built spinning
 probe with a maximum spinning frequency of about 10 kHz.


\section{$^{11}$B NMR in $\mbox{MgB}_2$ in the normal state}

The $^{11}$B NMR spectrum in MgB$_2$ powder samples is complicated
by the simultaneous presence of first and second order quadrupole
interactions, anisotropic Knight shift and a previously unnoticed
dipolar splitting which is particularly evident in $^{11}$B
isotopically enriched samples. In the following we analyze the
different spectral features, a necessary step in order to extract
reliable NMR parameters.

\subsection{Quadrupole interactions}

\begin{figure}
 \includegraphics[scale=1.14, draft=false]{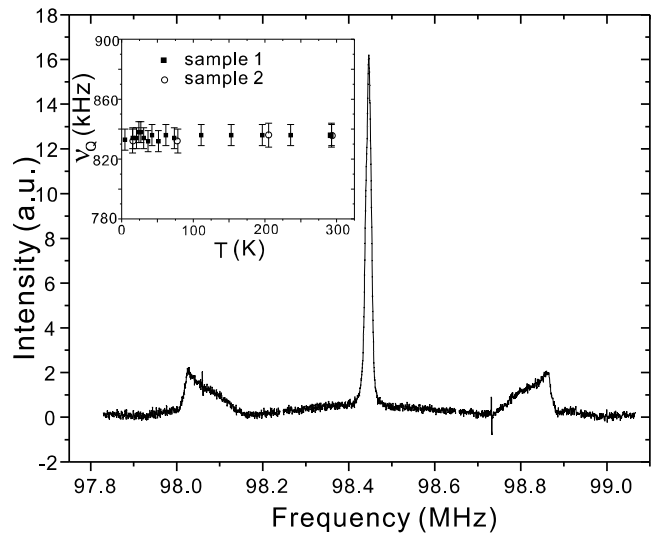}%
 \caption{\label{fig1}Room-temperature $^{11}$B NMR spectrum for MgB$_2$ powder sample
 showing both the central line transition and the two singularities of the distribution of satellite transitions.
 The inset shows the temperature dependence of the quadrupole coupling constant derived from
 the spectrum.}
 \end{figure}

The complete $^{11}$B NMR spectrum was determined in a 7.2 T
external field by performing the Fourier transform of half of the
solid echo following a $(\pi/2)_0-\tau-(\pi/2)_{90}$ pulse
sequence. In order to cover the whole spectrum three separate
spectra were recorded at resonance frequencies centered at the
three lines and added together. The result is shown in
Fig.~\ref{fig1}. Each rf pulse has a bandwidth of about 100 kHz.
Since the spectrum is the result of irradiation at only three
different frequencies, the parts of the spectrum in between the
peaks may be distorted but this does not affect the conclusions.
The separation of the symmetric satellite lines is due to first
and second order quadrupole interactions. For spin $I=3/2$, with
electric quadrupole moment $Q$ and for an axially symmetric
electric field gradient tensor (EFG) with maximum component $q$
the separation is~\cite{abragam}
\begin{equation}\label{eq1}
  \Delta\nu=\nu_Q(3\cos^2\theta-1)\quad\mbox{with } \nu_Q=\frac{e^2qQ}{2h}
\end{equation}
where $\theta$ is the angle between the principal axis of the EFG
tensor and the applied field $H_0$.

In a powder pattern the two singularities in the satellite
distribution occur at $\theta=90^\circ$. Thus from the spectrum in
Fig.~\ref{fig1} we obtain $\nu_Q = 835 \pm 10$ kHz. The quadrupole
coupling constant is almost temperature independent as shown in
the inset of Fig.~\ref{fig1}. A small increase in the quadrupole
coupling constant with decreasing temperature was observed in
$^{111}$Cd perturbed angular $\gamma\gamma$--correlation (TDPAC)
experiments~\cite{tsvya} and is consistent with the effect of
lattice vibrations.~\cite{borsa}

\begin{figure}
 \includegraphics[scale=0.68, draft=false]{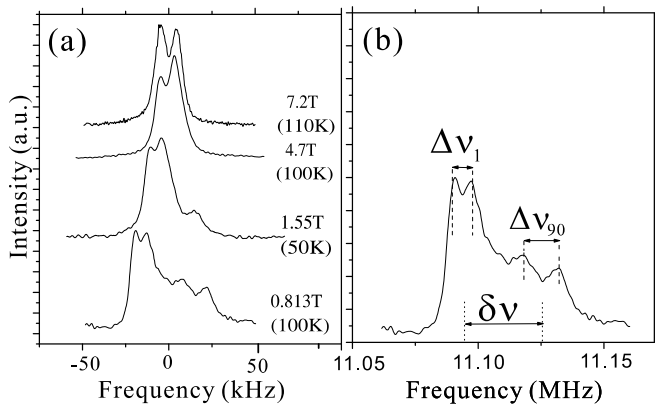}%
 \caption{\label{fig2}(a)~Central line transition of $^{11}$B NMR spectrum
 for powder sample of MgB$_2$ at several representative magnetic fields.
 (b)~Spectrum at 0.813~T. The distance between the middle point of each doublet
 corresponds to the second order quadrupole splitting, $\delta\nu$.}
 \end{figure}

In the presence of second order quadrupole effects the central
line transition is shifted. For axially symmetric EFG this results
in a powder pattern with two singularities separated by
\begin{equation}\label{eq2}
  \delta\nu =\frac{25}{48}\frac{\nu_Q^2}{\nu_L}
\end{equation}
where $\nu_L$ is the nuclear Larmor frequency. Because of the
inverse Larmor frequency dependence in Eq.~\ref{eq2} the two
singularities can be resolved only at low magnetic field as shown
in Fig.~\ref{fig2}(a).  At lower fields (1.55 and 0.813~T), the
second order quadrupole splitting is well resolved and each
singularity is split into a doublet. If one takes the middle point
of each doublet, the distance between singularities is in good
agreement with that calculated from Eq.~\ref{eq2} and the value of
$\nu_Q$ determined from the separation of the satellite lines i.e.
$\delta\nu = 32.7$ kHz at 0.813 T [see Fig.~\ref{fig2}(b)]. One
unexpected feature is that the two singularities are each split
into a field independent doublet.
 At higher fields (4.7 and 7.2~T), the second order quadrupole splitting is negligible
 and only one Pake doublet is shown.

\subsection{Dipolar doublet}
As shown in Fig.~\ref{fig2}(a) at high fields, where the second
order quadrupole effects become negligible, the $^{11}$B NMR
spectrum is formed by a doublet. At low fields one can clearly
resolve two doublets, one for each singularity in the quadrupole
powder pattern. The splitting is temperature and field independent
and results from the nearest neighbor nuclear dipolar interaction
of $^{11}$B nuclei in the planar honeycomb lattice structure. In
such a structure each $^{11}$B nucleus is strongly coupled to
three nearest neighbors resulting in three equivalent Pake
pairs.~\cite{pake} A single  dipolar pair gives rise to a Pake
doublet in the NMR spectrum with frequencies:
\begin{equation}\label{eq3}
  \nu_\pm = \nu_L \pm \nu_D(3\cos^2\theta'-1)
\end{equation}
with
\begin{equation}\label{eq3-1}
    \nu_D=\frac{3}{2}\frac{\gamma\mu}{2\pi a^3}
\end{equation}
where $\theta'$ is the angle formed by the magnetic field with the
vector joining the two interacting nuclei, $\mu$ is the nuclear
magnetic moment and $a$ is the internuclear distance.

In the presence of three interacting pairs one has to sum over the
different angles and take the powder average including the second
order quadrupole interaction and the anisotropic Knight shift
interaction. The situation was analyzed previously in connection
with measurements in intermetallic compounds of the C32 (AlB$_2$)
structure.~\cite{torgeson}  The conclusion was reached that in the
C32 structure the resonance line in the presence of all the above
interactions can be described in terms of the angle $\theta$
formed by the magnetic field and the $c$ axial symmetry axis
perpendicular to the B plane. Thus it was predicted that the
$\theta=90^\circ$ singularity in the quadrupole pattern should be
split into two lines separated by $\Delta\nu_{90}=3\nu_D$ while
the intermediate angle singularity should be split into two lines
separated by $\Delta\nu_{1}=2\nu_D$. From the inspection of
Fig.~\ref{fig2}(b) one can see that the ratio of the two
splittings is in reasonable agreement with the above prediction.
Furthermore, by using in Eq.~\ref{eq3-1} $\gamma= 13.66\times
2\pi$ MHz/T, $\mu= I \gamma h/2\pi$ and the measured value
$\nu_D=4.00\pm 0.50$ kHz (at 0.813~T), one obtains the B-B nearest
neighbor distance $a=1.72\pm 0.08$ \AA in agreement with the known
value of 1.782 \AA.

\subsection{Knight shift}

\begin{figure}
\includegraphics[scale=0.76, draft=false]{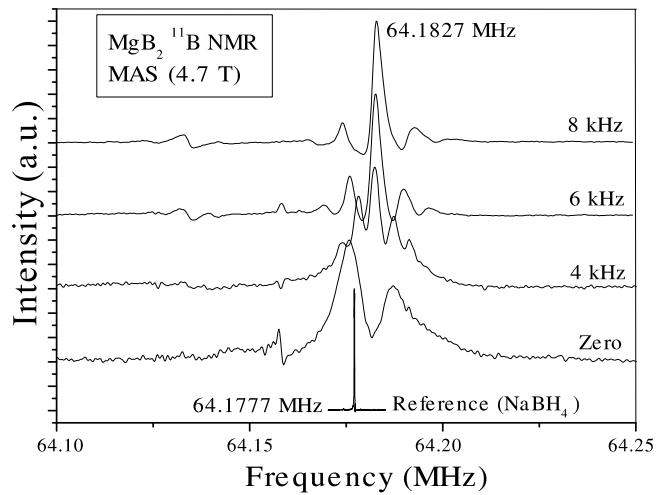}%
 \caption{\label{fig3}Magic angle spinning (MAS) experiment in
 MgB$_2$ powder sample at 4.7~T. Spinning frequencies are shown at the
 right side of the figure.}
 \end{figure}

In order to determine the Knight shift, both isotropic and
anisotropic parts, measurements should be made at very high
magnetic fields ($H \gg 10$ T) where the second order quadrupole
effects are negligible, and the anisotropic Knight shift may  be
inferred from the asymmetric broadening of the  NMR
line.~\cite{torgeson} Since our measurements are limited to a
maximum field of 7.2 T no measurable anisotropic Knight shift
could be detected. The isotropic Knight shift, $K$, is measurable
but very small. Its measurement is further complicated by the
dipolar splitting discussed above. In order to obtain a reliable
value of $K$ we performed a  magic angle spinning (MAS)
experiment~\cite{fukushima} at room temperature in a field of 4.7
T. As shown in Fig.~\ref{fig3}, the dipolar splitting is totally
removed for a spinning frequency of 8 kHz as well as part of the
dipolar broadening and of the quadrupole broadening. From the
resulting narrow line we obtain $K= 80$ ppm with respect to a
reference solution of NaBH$_4$.  If the Knight shift is referred
to the BF$_3$ solution, which is the compound used by chemists as
the ``zero chemical shift",~\cite{onak} one obtains $K= 40 \pm 10$
ppm.

\subsection{Nuclear spin-lattice relaxation rate}

\begin{figure}
\includegraphics[scale=0.95, draft=false]{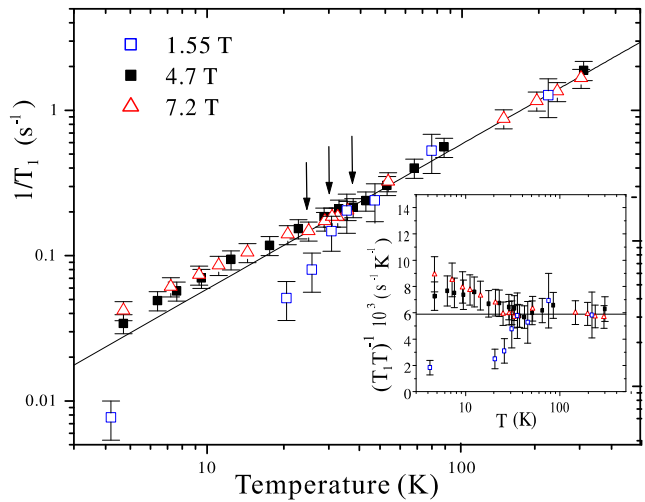}%
 \caption{\label{fig4}Temperature dependence of $1/T_1$ for $^{11}$B in MgB$_2$
 at external fields 1.55, 4.7
 and 7.2 T. The line is the Korringa law with $T_1T=170$ sK. The three
 arrows indicate the superconducting transition temperatures at each field
 which are 23, 27.5 and 34 K respectively.
 In the inset, $(T_1T)^{-1}\,10^3$ is plotted against temperature.}
 \end{figure}

For the case of saturation of the central line of the $^{11}$B NMR
spectrum with a single rf pulse (or short sequence of pulses) and
for magnetic relaxation mechanism, the recovery of the nuclear
magnetization after a time $t$ following  saturation is given
by:~\cite{andrew}
\begin{equation}\label{eq4}
  \frac{M(\infty)-M(t)}{M(\infty)}=0.1\exp(-2Wt)+0.9\exp(-12Wt)
\end{equation}
where we define  the nuclear spin lattice relaxation rate as
$1/T_1 = 2W$. The results for the temperature dependence of
$1/T_1$ are shown in Fig.~\ref{fig4}. As can be seen, a field
independent linear temperature dependence is observed in the
normal phase yielding $T_1T = 170$ sK.

\section{$^{11}$B and $^{27}\mbox{Al}$ NMR in $\mbox{AlB}_2$}

\begin{figure}
 \includegraphics[scale=0.67, draft=false]{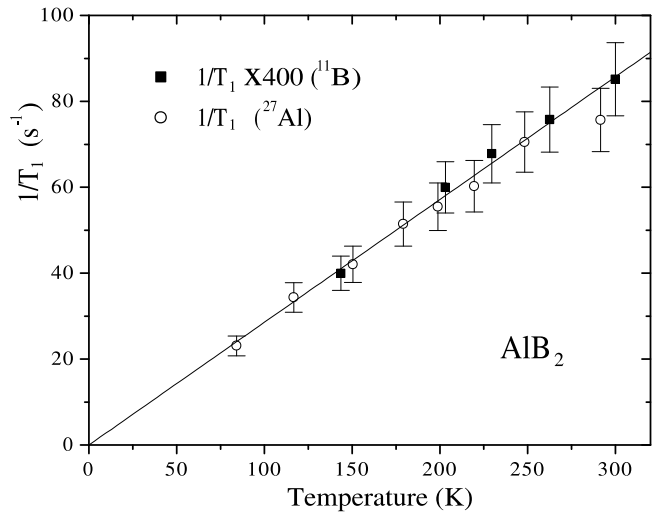}%
 \caption{\label{fig5}Comparison of $1/T_1$ for $^{11}$B  and for $^{27}$Al in AlB$_2$
 powder sample. $T_1T$ for $^{27}$Al is 400 times smaller than the
 value for $^{11}$B.}
 \end{figure}

The room temperature $^{11}$B NMR spectrum on the powder sample of
AlB$_2$ is very similar to the one shown in Fig.~\ref{fig1} for
MgB$_2$. From the separation of the satellite lines one derives a
quadrupole coupling frequency  $\nu_Q = 540 \pm 10$ kHz somewhat
smaller than the one in MgB$_2$. The  central transition linewidth
at room temperature is about 19 kHz and it hides the dipolar
splitting discussed for the case of MgB$_2$ due to the stronger
dipolar interaction with the $^{27}$Al nuclei present in AlB$_2$.
By spinning the sample at 10 kHz we obtain a MAS NMR line 2 kHz
wide. The Knight shift value with respect to a reference solution
of BF$_3$ measured from the MAS spectrum is found to be $K= -10
\pm 5$ ppm. Finally, the $1/T_1$ measurements as a function of
temperature shown in Fig.~\ref{fig5} yield a Korringa law with
$T_1T =1400$ sK i.e.~almost one order of magnitude greater than in
MgB$_2$.

The room temperature $^{27}$Al ($I=5/2$) NMR spectrum in AlB$_2$
is composed of a central transition line 23 kHz wide and a
poorly-resolved powder pattern originating from the two satellite
pairs over a spectral distribution of about 200 kHz. From the fit
of the NMR spectrum to a computer simulated spectrum with a
Gaussian dipolar width of  7.5 kHz  one can derive a quadrupole
coupling constant $\nu_Q = 80 \pm 10$ kHz. It is assumed that the
electric field gradient has axial symmetry as for the $^{11}$B
site. The Knight shift was measured from the position of the
central line transition (which has a negligible second order
quadrupole broadening). The Knight shift with respect to a
AlCl$_3$ aqueous solution is $K= 880 \pm 20$ ppm.

The $^{27}$Al $1/T_1$ results as a function of temperature are
shown in Fig.~\ref{fig5} together with the $^{11}$B results. The
linear temperature dependence yields for $^{27}$Al $T_1T=3.5$ sK
which is 400 times smaller than the value for $^{11}$B.

The various NMR parameters  for both MgB$_2$ and AlB$_2$ are
summarized in Tab.~\ref{table1}. The constant $S$ listed in
Tab.~\ref{table1} is given by $S= (\gamma_e/\gamma_n)^2 h/(8\pi^2
k_B)$ and the Korringa ratio $R$ is defined as $R= K^2 T_1T/S$.

\begin{table*}
\begin{tabular*}{7in}[b]{@{\extracolsep{\fill}}c|ccc|ccc}
\hline \hline
    & \multicolumn{2}{c} {MgB$_2$} & & \multicolumn{2}{c}      {AlB$_2$}                              \\
\hline
     & $^{25}$Mg [Ref.~\onlinecite{mali}] &    $^{11}$B   &  &    $^{27}$Al    &    $^{11}$B         &  \\
\hline \hline
     $K$ (ppm)   & $242\pm4$\footnotemark[1] &       +40 $\pm$ 10\footnotemark[2] &  & +880 $\pm$ 20\footnotemark[3]  & $-$10 $\pm$ 5\footnotemark[2]     & \\
    $T_1T$  (sK)  &   1090     &         170         & &         3.5          &         1400         \\
     $S$ (sK)  &   $7.03\times 10^{-5}$ & $2.57\times 10^{-6}$ & & $3.88\times 10^{-6}$ & $2.57\times 10^{-6}$ &\\
   $R$ & 0.95 &   0.102 $\pm$ 0.05 & & 0.7 $\pm$ 0.03 &  $0-0.12$       &  \\
 $\nu_Q$ (kHz) & 222 (1.5) & 835 (10) & & 80 (10) & 540 (10) &\\
 \hline\hline
\end{tabular*}
\caption{\label{table1}Summary of the various NMR parameters for
both MgB$_2$ and AlB$_2$. The Korringa ratio, $R$, is defined as
$R=K^2T_1T/S$. Results for $^{25}$Mg NMR were taken from
Ref.~[\onlinecite{mali}].} \footnotetext[1]{Reference solution
MgCl$_2$} \footnotetext[2]{Reference solution BF$_3$}
\footnotetext[3]{Reference solution AlCl$_3$}
\end{table*}

\section{Comparison of theory and experiments in $\mbox{MgB}_2$ and $\mbox{AlB}_2$}

In order to understand the microscopic origin of the relaxation,
and thus also the differences between AlB$_2$ and MgB$_2$, a
comparison between experimental and {\em ab-initio} calculated
values of Knight shifts and relaxation rates is highly desirable.
Recently, first principles calculation of the relaxation
rates~\cite{pavarini,prbant} and the Knight shifts~\cite{pavarini}
were performed for MgB$_2$. In Ref.~[\onlinecite{prbant}] the
relaxation rates were calculated for AlB$_2$ as well. In the
present work we calculate $K$ and $1/T_1T$ for AlB$_2$ by using
the method described in Ref.~[\onlinecite{pavarini}], and compare
the results for MgB$_2$ and AlB$_2$. Our calculations are based on
density functional theory (DFT) in the local density approximation
(LDA). We adopt the tight binding linear-muffin-tin-orbital (LMTO)
method~\cite{lmto} in the atomic-spheres-approximation (LMTO47
Stuttgart code). The  density of states matrix (Eq.~2 of
Ref.~[\onlinecite{pavarini}]) and the partial density of states,
$N_L$, (with $L=lm$, where $l$ is the orbital angular momentum
quantum number, and $m=-l,\cdots,l$) were calculated by using the
linear tetrahedron method. We found that the results already
converged very well with a mesh of 481 irreducible {\bf k} points.
In order to obtain accurate wavefunctions at the Fermi level, the
linear partial wave expansion was performed with
$\epsilon_\nu\equiv\epsilon_F$, where $\epsilon_F$ is the Fermi
level and $\epsilon_\nu$ the expansion energy. Further details
about the method employed can be found in
Ref.~[\onlinecite{pavarini}].

In Tab.~\ref{table2} and Tab.~\ref{table3}, we show the calculated
$K$ and $1/T_1T$. The results for MgB$_2$ are taken from
Ref.~[\onlinecite{pavarini}].

Let us discuss first the case of MgB$_2$. The $^{11}$B orbital
relaxation rate is about 3 times larger than the dipole-dipole,
and about 10 times larger than the Fermi-contact term. As
explained in Ref.~[\onlinecite{pavarini}], the reason is the
following. In MgB$_2$ the B $p_\sigma$ and B $p_\pi$ bands are all
at the Fermi level: the partial density of states $N_{p_{x,y}}$
($\sigma$ bands) and $N_{p_z}$ ($\pi$ bands) are, respectively,
$N_{p_x}=N_{p_y}\sim0.035$ states/eV/spin/B, and
$N_{p_z}\sim0.045$ states/eV/spin/B. On the contrary, there are
very few B $s$ electrons close to $\epsilon_F$ ($N_s\sim 0.002$
states/eV/B). An approximate formula for the ratio between the
Fermi-contact and the orbital/dipole-dipole coupling constants is
given by~\cite{pavarini}
\begin{equation}
F={2\over 3} {|\phi_s(0)|^2 N_s \over
 \sum_{l>0} \langle r^{-3}\rangle_{ll} N_l}.
\end{equation}
Here $\phi_l(r)$ is the $l$ radial solutions of the Schr\"odinger
equation at energy $\epsilon_\nu$ and $N_l\equiv\sum_{m=-l}^{m=l}
N_{lm}$. In addition $\langle r^{-3}\rangle_{ll}\equiv \int
(|\phi_l(r)|^2 / r^{-3}) r^2 dr $ and
$\phi_s(0)\equiv\phi_{l=0}(r=0)$, where $r=0$ is the position of
the nucleus. In the case of B it was found~\cite{pavarini} $F\sim
0.35$. Thus $F$ is considerably smaller than 1, and the Fermi
contact interaction (which usually dominates over the orbital and
dipole-dipole mechanisms) gives only a small contribution to the
relaxation rate. Moreover, it was found~\cite{pavarini}  that
$N_{p_x}= N_{p_y}\sim N_{p_z}$. For a model Hamiltonian which
includes only B $p$ orbitals and for which
$N_{p_x}\!=\!N_{p_y}\!=\!N_{p_z}\!=\! N_p/3$, one can show
analytically~\cite{pavarini} that the ratio between
orbital/dipole-dipole relaxation rate is about 3.3. This explains
the numerical results obtained for B in MgB$_2$.

The case of $^{25}$Mg is different: the ratio $F$ is considerably
larger than one ($F\sim 5$) and thus the Fermi contact interaction
is expected to dominate. Tab.~\ref{table3} shows that this
actually happens: the orbital and the dipole-dipole terms are much
smaller than the Fermi contact contribution.

With a Stoner enhancement factor of about $1.33$ (calculated
ab-initio in Ref.~[\onlinecite{pavarini}]), the following results
were found for the total Knight shifts: $K$(Mg)$\;\sim 361/341$
ppm and $K$(B)$\;\sim 21/37$ ppm. For the relaxation time, it was
shown~\cite{pavarini} that the Stoner enhancement factor is about
1.6. Thus the total relaxation times are:  $T_1T$(Mg)$\;\sim 625 $
sK and $T_1T$(B)$\;\sim 169$ sK. These results are in quite good
agreement with experimental data.

Let us now discuss the case of AlB$_2$. In this compound the B
$p_\sigma$ bands are fully occupied, and only B $p_\pi$ bands are
at the Fermi level. In addition  $|\phi_s(0)|^2/4\pi \sim 2.87
\,a_0^3$ and $ N_s \sim 0.003$  states/eV/spin/B, $\langle
r^{-3}\rangle_{11} \sim 1.45\, a_0^3$ and $N_p \sim 0.0216$
states/eV/spin/B. With these numbers we find $F\sim 2.3$; terms
with $l>1$ give a small contribution to $F$, because the radial
integrals decrease quickly when $l$ increases and because, in the
case of B, $N_l$ is very small for $l>1$. Thus $F$ is considerably
larger than 1, and the Fermi contact interaction is the dominant
mechanism of relaxation for $^{11}$B (see Tab.~\ref{table3}). The
same happens for $^{27}$Al, for which we find
$|\phi_s(0)|^2/4\pi\sim 2.96 \,a_0^3$  and $N_s\sim0.0362$
states/eV/spin, $\langle r^{-3}\rangle_{11}\sim1.74 \,a_0^3$ and
$N_p\sim0.0325$ states/eV/spin/B, and therefore $F\sim 16$.

\begin{table*}
\center {\begin{tabular*}{7in}[b]{@{\extracolsep{\fill}}ccccccccc}
\hline
MgB$_2$& dipole (xy) & dipole(z) & orbital&  Fermi contact& core & Total (xy/z) & Experiment \\
\hline
B  &-4 & 8 & 0& 27  &-7 &16/28 & $40\pm 10$\\
Mg & 5 &-10 &0&260 & 3 &271/256 & $242\pm 4$\\
\hline
AlB$_2$& dipole (xy) & dipole(z) & orbital &  Fermi contact& core & Total (xy/z) & Experiment\\
\hline
B  &-8  & 16 & 0& 61   &-11 &  42/66 & $-10\pm 5$  \\
Al & 1  &-2  & 0& 660  &-15 & 644/647& $880\pm 20$ \\
\hline
\end{tabular*}}
\caption[short]{\label{table2} Knight shifts in ppm in MgB$_2$ and
AlB$_2$ (without Stoner factor).
 The label $\alpha=x,y,z$ indicates the direction of the
 external magnetic field. Results for MgB$_2$ are
 taken from Ref.~[\onlinecite{pavarini}].}
\end{table*}

\begin{table*}
\center {\begin{tabular*}{7in}[b]{@{\extracolsep{\fill}}cccccccc}
\hline
MgB$_2$& dipole & orbital &  Fermi-contact& core & Total &Experiment \\
\hline
B  &0.8 & 2.6  & 0.28 &0.02   &3.7 & 5.9\\
Mg &0.01 & 0.02 & 1.0  &0.0001 &1.0 & 0.92\\
\hline
AlB$_2$& dipole & orbital &  Fermi-contact& core & Total & Experiment\\
\hline
B  &0.086    &0.132 & 1.4 &  0.04 &  1.66 & 0.71\\
Al &0.115    &0.370 & 105 &  0.05  &  105 & 286\\
\hline
\end{tabular*}}
\caption[short]{\label{table3} Relaxation rates in $10^{-3}/$(sK)
in MgB$_2$ and AlB$_2$ (without Stoner factor).
 Results for MgB$_2$ are taken from Ref.~[\onlinecite{pavarini}].}
\end{table*}

In order to understand better the numerical results for AlB$_2$,
we calculate analytically the contact shift and relaxation rates
for a model Hamiltonian which includes only B $s$ and Al $s$
states. Within this model, the Knight shift is given by $K\sim
\mu_B^2 (4/3) |\phi_s(0)|^2 N_s$, and the relaxation rate can be
obtained from the Korringa relation. We find $K$(Al)$\;\sim 645$
ppm, $K$(B)$\;\sim 52$ ppm and $1/T_1T$(Al)$\;\sim 107\cdot
10^{-3}/$(sK), $1/T_1T$(B)$\;\sim 1.05\cdot 10^{-3}/$(sK), in very
good agreement with the first principles values of
Tab.~\ref{table2} and Tab.~\ref{table3}.

The {\em ab-initio} total Knight shifts (Tab.~\ref{table2}) are
$K$(Al)$\;\sim644/647$ ppm and $K$(B)$\;\sim42/66$, and the total
relaxation times (Tab.~\ref{table3}) are $T_1T$(Al)$\;\sim 9$ sK
and $T_1T$(B)$\;\sim 602 $ sK. The agreement between first
principle results and  experimental data is quite good for Al. In
the case of B the calculated relaxation time is about 2 times
smaller than the experimental data. Similar results were found in
Ref.~[\onlinecite{prbant}]. This discrepancy could suggest that,
in the case of AlB$_2$, LDA tends to slightly overestimate the
Fermi-contact interaction at B nucleus.

Finally, we point out that in AlB$_2$ the density of states in the
B plane is strongly reduced with respect to MgB$_2$. The reason is
that in AlB$_2$ the $\sigma$ bands are well below $\epsilon_F$,
while in MgB$_2$ they cross the Fermi level. We find
$N_{p_x}+N_{p_y}\sim 0.0046$ and $N_{p_z}\sim 0.017$
states/eV/spin/B, while for MgB$_2$ $N_{p_x}+N_{p_y}\sim 0.07$ and
$N_{p_z}\sim 0.045$ states/eV/spin/B was found.~\cite{pavarini}

\section{$^{11}$B  NMR in $\mbox{MgB}_2$ in the superconducting phase}

Although the main emphasis of the present paper is on the
electronic properties of MgB$_2$ in the normal phase, it is
worthwhile to present and discuss here the $^{11}$B NMR results
obtained in the superconducting phase mostly to point out the
limitations incurred in the NMR experiments in polycrystalline
samples.

Regarding the Knight shift, one can conclude that no meaningful
measurements of $K$ can be performed below $T_c$ in a powder
sample. In fact the Knight shift is very small compared to the
line broadening due to the magnetic field distribution of the flux
line lattice.~\cite{papa}  Furthermore  the correction for the
shift due to the diamagnetic shielding in the superconducting
phase cannot be estimated accurately in a powder sample as a
result of the distribution of shapes of the particles
e.g.~distribution of demagnetization factors.

  \begin{figure}
\includegraphics[scale=0.65, draft=false]{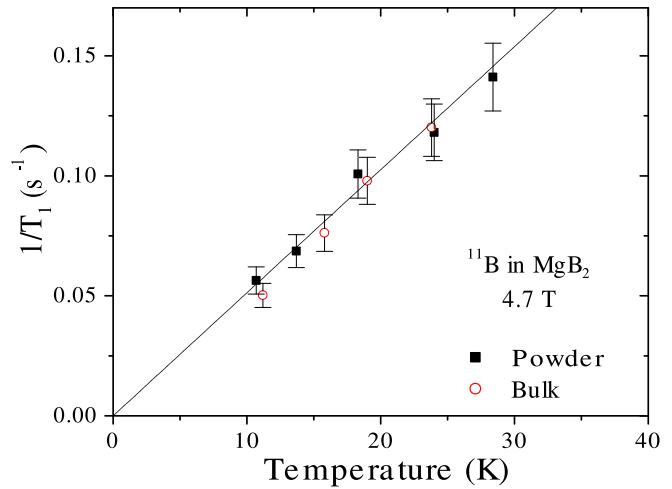}%
 \caption{\label{fig6}$1/T_1$ results for $^{11}$B in a powder and in a sintered
 bulk sample of MgB$_2$ below $T_c$.}
 \end{figure}

The temperature dependence of $1/T_1$ below $T_c$ allows in
principle to obtain information about the pairing symmetry and the
structure of the superconducting gap.~\cite{hebel,hebel2} In fact
the ratio of the relaxation rates in the superconducting phase and
the normal phase, $(1/T_{1s})/(1/T_{1n})$, is related to the
density of states in the superconducting phase and should decrease
below $T_c$ either exponentially or with a power law depending on
the pairing symmetry and/or the structure of the
gap.~\cite{kotegawa} As can be seen in Fig.~\ref{fig4} a decrease
of the relaxation rate below $T_c$ can be observed in the data
taken in an external field of 1.55 T but not in the data at 4.7 T
and 7.2 T. The explanation for this is easily found in the strong
anisotropy of the upper critical field, $H_{c2}$. The powders
utilized in the present experiment have $H_{c2}^{max}/H_{c2}^{min}
= \gamma \approx 6$ whereby the maximum critical field pertains to
the particles with the field in the $ab$ plane.~\cite{budko,simon}
In a detuning experiment one detects the superconducting
transition of the particles which have the higher upper critical
field (see arrows in Fig.~\ref{fig4}). On the other hand in the
NMR experiment the stronger signal arises from the particles which
are oriented in such a way as to have the lower upper critical
field. This is a consequence of the strongly reduced rf
penetration length and consequently reduced NMR signal, in the
superconducting particles. Thus the results in Fig.~\ref{fig4} for
fields of 4.7~T and 7.2~T pertain mostly to the particles which
remain in the normal phase down to helium temperature. It should
be noted that the results in a loose powder and in a
polycrystalline bulk (sintered) sample are the same as shown in
Fig.~\ref{fig6}.

  \begin{figure}
\includegraphics[scale=0.57, draft=false]{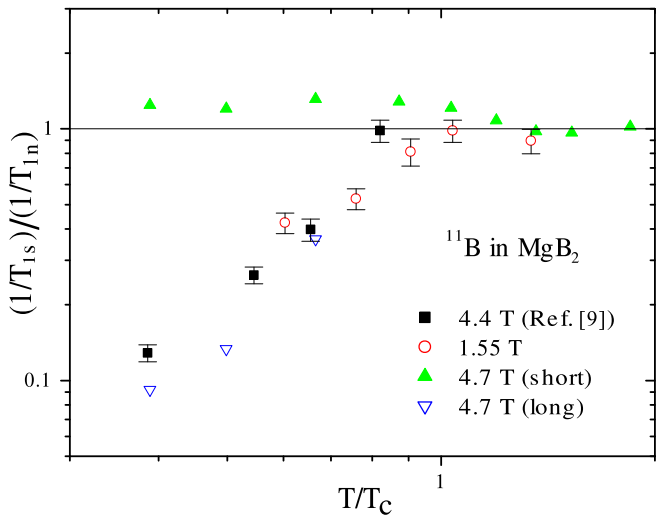}%
 \caption{\label{fig7}The plot of the ratio of $1/T_1$ in the superconducting
 phase and the normal phase against $T/T_c$. The short and long components of the relaxation rate
 at 4.7~T were obtained by fitting the data to
 Eq.~\ref{eq4} with two different values of $W$.}
 \end{figure}

Our results need to be reconciled with the $^{11}$B NMR data in
MgB$_2$ by Kotegawa et al.~\cite{kotegawa} where it is reported
that $1/T_1$ decreases exponentially below $T_c$ in powder samples
for all field values from 1.35~T up to 4.42~T  with a tiny
coherence peak just below $T_c$. First it is noted  that the
samples used in Kotegawa's experiments could have a much smaller
anisotropy of the upper critical field. Recent resistivity
measurements in MgB$_2$ single crystals yield an anisotropy ratio
of the upper critical field of only 2.7.~\cite{lee} A second
factor is that our measurements were done by fitting the first 90
\% of the recovery of the nuclear magnetization which yields the
short component only of the relaxation. On the other hand the
results reported by Kotegawa et al.~were obtained by fitting the
recovery of the magnetization with two components and assuming
that the long component only is the one pertaining to the
superconducting particles.~\cite{kitaoka} By following the same
procedure i.e.~by fitting the recovery of the nuclear
magnetization with Eq.~\ref{eq4} with two different values of $W$
we find the results shown in Fig.~\ref{fig7}. The long component
does indeed agree with Kotegawa's data at high field (4.7~T) and
with our low field data (1.55~T)  thus confirming the validity of
the conclusions regarding the structure of the
gap.~\cite{kotegawa} However, the distribution of critical fields
due to the random orientation of the particles in a powder sample
introduces large errors in the evaluation of $1/T_1$ particularly
close to $T_c$ since the recovery of the magnetization is the
superposition of curves, Eq.~\ref{eq4}, with different $W$ values.
Therefore it is very difficult to infer the presence of a
coherence peak in $1/T_1$ just below $T_c$ as claimed by Kotegawa
et al.~\cite{kotegawa} It is noted that from the data in the inset
of Fig.~\ref{fig4} and Fig.~\ref{fig7} there appears to be a
enhancement of $(T_1T)^{-1}$ over the Korringa value of the normal
phase starting at $T_c$ and extending to low temperature. This
effect, which is barely outside the experimental error, is not
presently understood. One may speculate that the measured $1/T_1$
in Fig.~\ref{fig4} is an average of the relaxation of nuclei in
particles in the normal phase and particles in the superconducting
phase with $H$ perpendicular to the $ab$ plane in which the
relaxation is enhanced by the presence of flux
lines.~\cite{rigamonti} More detailed measurements as a function
of magnetic field are needed to elucidate this point.

\section{Summary and conclusions}

From the analysis of the $^{11}$B NMR spectrum in the normal phase
of isotopically enriched MgB$_2$, we have found evidence for a
field independent splitting of the line due to the nuclear dipolar
interaction (Pake doublet). By averaging out the dipolar
interaction with Magic Angle Spinning, we have obtained reliable
values for the Knight shifts for $^{11}$B in both MgB$_2$ and
AlB$_2$. Both the decrease of $K$ and more so of $(T_1T)^{-1}$ for
$^{11}$B in AlB$_2$ with respect to MgB$_2$ is in qualitative
agreement with a drastic drop of the DOS at the Fermi level at the
B site in AlB$_2$. The ab-initio calculated values in
Tab.~\ref{table2} and~\ref{table3} are in excellent agreement with
the experimental ones in MgB$_2$ provided the theoretical results
are multiplied by a Stoner enhancement factor of 1.33 for the
Knight shift and 1.6 for the relaxation rates.

For the case of AlB$_2$ the ab-initio calculated values are in
good agreement with the experiments only for the $^{27}$Al data
considering that the theoretical results in Tab.~\ref{table2}
and~\ref{table3} are still to be multiplied by the Stoner factor.
However, for the B site in AlB$_2$ the theoretical values for both
$K$ and $(T_1T)^{-1}$ are a factor of two bigger than the
experimental values indicating that the LDA calculations
overestimate the Fermi contact interaction at the B nuclear site.

Regarding the superconducting phase in MgB$_2$, we conclude that
it is very difficult to obtain reliable information for the NMR
parameters (particularly the relaxation rates) from
polycrystalline samples due to the random orientation with respect
to the applied field and the strong critical field anisotropy. In
particular, we could confirm qualitatively the exponential drop of
$1/T_1$ vs $T$ below $T_c$ reported by Kotegawa et
al.~\cite{kotegawa} but we could not detect any enhancement of
$1/T_1$ just below $T_c$ (coherence peak).

One issue which remains to be investigated is the effect on the
nuclear relaxation rate of the flux line lattice and flux line
fluctuations in the superconducting phase which was detected
qualitatively in the enhancement of $T_1T$ below $T_c$ (see the
inset of Fig.~\ref{fig4}) but which would require aligned powder
samples or single crystals to be investigated quantitatively.

\begin{acknowledgments}
We thank J.~K.~Jung for his help in the beginning stage of the
experiment, P.~Carretta and Tabak for sharing some unpublished
data on different samples and V.~Antropov for useful discussions.
 Ames Laboratory is
operated for the U.S. Department of Energy by Iowa State
University under Contract W-7405-Eng-82. This work at Ames
Laboratory was supported by the Director for Energy Research,
Office of Basic Energy Science. One of authors (B.J.S.)
acknowledges the support by KOSEF via electron Spin Science Center
at POSTECH.
\end{acknowledgments}

\bibliography{nmr_mgb2}

\begin{thebibliography}{29}
\expandafter\ifx\csname natexlab\endcsname\relax\def\natexlab#1{#1}\fi
\expandafter\ifx\csname bibnamefont\endcsname\relax
  \def\bibnamefont#1{#1}\fi
\expandafter\ifx\csname bibfnamefont\endcsname\relax
  \def\bibfnamefont#1{#1}\fi
\expandafter\ifx\csname citenamefont\endcsname\relax
  \def\citenamefont#1{#1}\fi
\expandafter\ifx\csname url\endcsname\relax
  \def\url#1{\texttt{#1}}\fi
\expandafter\ifx\csname urlprefix\endcsname\relax\def\urlprefix{URL }\fi
\providecommand{\bibinfo}[2]{#2}
\providecommand{\eprint}[2][]{\url{#2}}

\bibitem[{\citenamefont{Nagamatsu et~al.}(2001)\citenamefont{Nagamatsu,
  Nakagawa, Muranaka, Zenitani, and Akimitsu}}]{naga}
\bibinfo{author}{\bibfnamefont{J.}~\bibnamefont{Nagamatsu}},
  \bibinfo{author}{\bibfnamefont{N.}~\bibnamefont{Nakagawa}},
  \bibinfo{author}{\bibfnamefont{T.}~\bibnamefont{Muranaka}},
  \bibinfo{author}{\bibfnamefont{Y.}~\bibnamefont{Zenitani}}, \bibnamefont{and}
  \bibinfo{author}{\bibfnamefont{J.}~\bibnamefont{Akimitsu}},
  \bibinfo{journal}{Nature} \textbf{\bibinfo{volume}{410}}, \bibinfo{pages}{63}
  (\bibinfo{year}{2001}).

\bibitem[{\citenamefont{Bud'ko et~al.}(2001{\natexlab{a}})\citenamefont{Bud'ko,
  Lapertot, Petrovic, Cunningham, Anderson, and Canfield}}]{budko2}
\bibinfo{author}{\bibfnamefont{S.~L.} \bibnamefont{Bud'ko}},
  \bibinfo{author}{\bibfnamefont{G.}~\bibnamefont{Lapertot}},
  \bibinfo{author}{\bibfnamefont{C.}~\bibnamefont{Petrovic}},
  \bibinfo{author}{\bibfnamefont{C.~E.} \bibnamefont{Cunningham}},
  \bibinfo{author}{\bibfnamefont{N.}~\bibnamefont{Anderson}}, \bibnamefont{and}
  \bibinfo{author}{\bibfnamefont{P.~C.} \bibnamefont{Canfield}},
  \bibinfo{journal}{Phys.\ Rev.\ Lett.} \textbf{\bibinfo{volume}{86}},
  \bibinfo{pages}{1877} (\bibinfo{year}{2001}{\natexlab{a}}).

\bibitem[{\citenamefont{Hinks et~al.}(2001)\citenamefont{Hinks, Claus, and
  Jorgensen}}]{hinks}
\bibinfo{author}{\bibfnamefont{D.~G.} \bibnamefont{Hinks}},
  \bibinfo{author}{\bibfnamefont{H.}~\bibnamefont{Claus}}, \bibnamefont{and}
  \bibinfo{author}{\bibfnamefont{J.~D.} \bibnamefont{Jorgensen}},
  \bibinfo{journal}{Nature} \textbf{\bibinfo{volume}{411}},
  \bibinfo{pages}{457} (\bibinfo{year}{2001}).

\bibitem[{\citenamefont{Rigamonti et~al.}(1998)\citenamefont{Rigamonti, Borsa,
  and Carretta}}]{rigamonti}
\bibinfo{author}{\bibfnamefont{A.}~\bibnamefont{Rigamonti}},
  \bibinfo{author}{\bibfnamefont{F.}~\bibnamefont{Borsa}}, \bibnamefont{and}
  \bibinfo{author}{\bibfnamefont{P.}~\bibnamefont{Carretta}},
  \bibinfo{journal}{Rep.\ Prog.\ Phys.} \textbf{\bibinfo{volume}{61}},
  \bibinfo{pages}{1367} (\bibinfo{year}{1998}).

\bibitem[{\citenamefont{Jung et~al.}(2001)\citenamefont{Jung, Baek, Borsa,
  Bud'ko, Lapertot, and Canfield}}]{jung}
\bibinfo{author}{\bibfnamefont{J.~K.} \bibnamefont{Jung}},
  \bibinfo{author}{\bibfnamefont{S.~H.} \bibnamefont{Baek}},
  \bibinfo{author}{\bibfnamefont{F.}~\bibnamefont{Borsa}},
  \bibinfo{author}{\bibfnamefont{S.~L.} \bibnamefont{Bud'ko}},
  \bibinfo{author}{\bibfnamefont{G.}~\bibnamefont{Lapertot}}, \bibnamefont{and}
  \bibinfo{author}{\bibfnamefont{P.~C.} \bibnamefont{Canfield}},
  \bibinfo{journal}{Phys.\ Rev.\ B} \textbf{\bibinfo{volume}{64}},
  \bibinfo{pages}{012514} (\bibinfo{year}{2001}).

\bibitem[{\citenamefont{Tsvyashchenko et~al.}(2001)}]{tsvya}
\bibinfo{author}{\bibfnamefont{A.~V.} \bibnamefont{Tsvyashchenko}}
  \bibnamefont{et~al.}, \bibinfo{journal}{Solid State Comm.}
  \textbf{\bibinfo{volume}{119}}, \bibinfo{pages}{153} (\bibinfo{year}{2001}).

\bibitem[{\citenamefont{Mali et~al.}(2001)\citenamefont{Mali, Roos, Shengelaya,
  Keller, and Conder}}]{mali}
\bibinfo{author}{\bibfnamefont{M.}~\bibnamefont{Mali}},
  \bibinfo{author}{\bibfnamefont{J.}~\bibnamefont{Roos}},
  \bibinfo{author}{\bibfnamefont{A.}~\bibnamefont{Shengelaya}},
  \bibinfo{author}{\bibfnamefont{H.}~\bibnamefont{Keller}}, \bibnamefont{and}
  \bibinfo{author}{\bibfnamefont{K.}~\bibnamefont{Conder}},
  \bibinfo{journal}{Cond-Mat/0111022 v2}  (\bibinfo{year}{2001}).

\bibitem[{\citenamefont{Papavassiliou et~al.}(2001)\citenamefont{Papavassiliou,
  Pissas, Fardis, Karayanni, and Christides}}]{papa}
\bibinfo{author}{\bibfnamefont{G.}~\bibnamefont{Papavassiliou}},
  \bibinfo{author}{\bibfnamefont{M.}~\bibnamefont{Pissas}},
  \bibinfo{author}{\bibfnamefont{M.}~\bibnamefont{Fardis}},
  \bibinfo{author}{\bibfnamefont{M.}~\bibnamefont{Karayanni}},
  \bibnamefont{and}
  \bibinfo{author}{\bibfnamefont{C.}~\bibnamefont{Christides}},
  \bibinfo{journal}{Cond-Mat/0107511}  (\bibinfo{year}{2001}).

\bibitem[{\citenamefont{Kotegawa et~al.}(2001)\citenamefont{Kotegawa, Ishida,
  Kitaoka, Muranaka, and Akimitsu}}]{kotegawa}
\bibinfo{author}{\bibfnamefont{H.}~\bibnamefont{Kotegawa}},
  \bibinfo{author}{\bibfnamefont{K.}~\bibnamefont{Ishida}},
  \bibinfo{author}{\bibfnamefont{Y.}~\bibnamefont{Kitaoka}},
  \bibinfo{author}{\bibfnamefont{T.}~\bibnamefont{Muranaka}}, \bibnamefont{and}
  \bibinfo{author}{\bibfnamefont{J.}~\bibnamefont{Akimitsu}},
  \bibinfo{journal}{Phys.\ Rev.\ Lett.} \textbf{\bibinfo{volume}{87}},
  \bibinfo{pages}{127001} (\bibinfo{year}{2001}).

\bibitem[{\citenamefont{Gerashenko et~al.}(2001)\citenamefont{Gerashenko,
  Mikalhev, Verkhovskij, D'yachova, Tyutyunnik, and Zubkov}}]{gera}
\bibinfo{author}{\bibfnamefont{A.}~\bibnamefont{Gerashenko}},
  \bibinfo{author}{\bibfnamefont{K.}~\bibnamefont{Mikalhev}},
  \bibinfo{author}{\bibfnamefont{S.}~\bibnamefont{Verkhovskij}},
  \bibinfo{author}{\bibfnamefont{T.}~\bibnamefont{D'yachova}},
  \bibinfo{author}{\bibfnamefont{A.}~\bibnamefont{Tyutyunnik}},
  \bibnamefont{and} \bibinfo{author}{\bibfnamefont{V.}~\bibnamefont{Zubkov}},
  \bibinfo{journal}{Cond-Mat/0102421}  (\bibinfo{year}{2001}).

\bibitem[{\citenamefont{Tou et~al.}(2001)\citenamefont{Tou, Ikejiri, Maniwa,
  Ito, Takenobu, Prassides, and Iwasa}}]{tou}
\bibinfo{author}{\bibfnamefont{H.}~\bibnamefont{Tou}},
  \bibinfo{author}{\bibfnamefont{H.}~\bibnamefont{Ikejiri}},
  \bibinfo{author}{\bibfnamefont{Y.}~\bibnamefont{Maniwa}},
  \bibinfo{author}{\bibfnamefont{T.}~\bibnamefont{Ito}},
  \bibinfo{author}{\bibfnamefont{T.}~\bibnamefont{Takenobu}},
  \bibinfo{author}{\bibfnamefont{K.}~\bibnamefont{Prassides}},
  \bibnamefont{and} \bibinfo{author}{\bibfnamefont{Y.}~\bibnamefont{Iwasa}},
  \bibinfo{journal}{Cond-Mat/0103484}  (\bibinfo{year}{2001}).

\bibitem[{\citenamefont{Finnemore et~al.}(2001)\citenamefont{Finnemore,
  Ostenson, Bud'ko, Lapertot, and Canfield}}]{finnemore}
\bibinfo{author}{\bibfnamefont{D.~K.} \bibnamefont{Finnemore}},
  \bibinfo{author}{\bibfnamefont{J.~E.} \bibnamefont{Ostenson}},
  \bibinfo{author}{\bibfnamefont{S.~L.} \bibnamefont{Bud'ko}},
  \bibinfo{author}{\bibfnamefont{G.}~\bibnamefont{Lapertot}}, \bibnamefont{and}
  \bibinfo{author}{\bibfnamefont{P.~C.} \bibnamefont{Canfield}},
  \bibinfo{journal}{Phys.\ Rev.\ Lett} \textbf{\bibinfo{volume}{86}},
  \bibinfo{pages}{2420} (\bibinfo{year}{2001}).

\bibitem[{\citenamefont{Canfield et~al.}(2001)\citenamefont{Canfield,
  Finnemore, Bud'ko, Ostenson, Lapertot, Cunningham, and Petrovic}}]{canfield}
\bibinfo{author}{\bibfnamefont{P.~C.} \bibnamefont{Canfield}},
  \bibinfo{author}{\bibfnamefont{D.~K.} \bibnamefont{Finnemore}},
  \bibinfo{author}{\bibfnamefont{S.~L.} \bibnamefont{Bud'ko}},
  \bibinfo{author}{\bibfnamefont{J.~E.} \bibnamefont{Ostenson}},
  \bibinfo{author}{\bibfnamefont{G.}~\bibnamefont{Lapertot}},
  \bibinfo{author}{\bibfnamefont{C.~E.} \bibnamefont{Cunningham}},
  \bibnamefont{and} \bibinfo{author}{\bibfnamefont{C.}~\bibnamefont{Petrovic}},
  \bibinfo{journal}{Phys.\ Rev.\ Lett.} \textbf{\bibinfo{volume}{86}},
  \bibinfo{pages}{2423} (\bibinfo{year}{2001}).

\bibitem[{\citenamefont{Abragam}(1961)}]{abragam}
\bibinfo{author}{\bibfnamefont{A.}~\bibnamefont{Abragam}},
  \emph{\bibinfo{title}{Principles of Nuclear Magnetism}}
  (\bibinfo{publisher}{Academic Press, London}, \bibinfo{year}{1961}).

\bibitem[{\citenamefont{Borsa and Rigamonti}(1975)}]{borsa}
\bibinfo{author}{\bibfnamefont{F.}~\bibnamefont{Borsa}} \bibnamefont{and}
  \bibinfo{author}{\bibfnamefont{A.}~\bibnamefont{Rigamonti}},
  \bibinfo{journal}{J.\ of Magnetic Resonance} \textbf{\bibinfo{volume}{20}},
  \bibinfo{pages}{232} (\bibinfo{year}{1975}).

\bibitem[{\citenamefont{Pake}(1948)}]{pake}
\bibinfo{author}{\bibfnamefont{G.~E.} \bibnamefont{Pake}},
  \bibinfo{journal}{J.\ Chem.\ Phys.} \textbf{\bibinfo{volume}{16}},
  \bibinfo{pages}{327} (\bibinfo{year}{1948}).

\bibitem[{\citenamefont{Torgeson et~al.}(1972)\citenamefont{Torgeson, Barnes,
  and Creel}}]{torgeson}
\bibinfo{author}{\bibfnamefont{D.~R.} \bibnamefont{Torgeson}},
  \bibinfo{author}{\bibfnamefont{R.~G.} \bibnamefont{Barnes}},
  \bibnamefont{and} \bibinfo{author}{\bibfnamefont{R.~B.} \bibnamefont{Creel}},
  \bibinfo{journal}{J.\ Chem.\ Phys.} \textbf{\bibinfo{volume}{56}},
  \bibinfo{pages}{4178} (\bibinfo{year}{1972}).

\bibitem[{\citenamefont{Fukushima and Roeder}(1981)}]{fukushima}
\bibinfo{author}{\bibfnamefont{E.}~\bibnamefont{Fukushima}} \bibnamefont{and}
  \bibinfo{author}{\bibfnamefont{S.~B.~W.} \bibnamefont{Roeder}},
  \emph{\bibinfo{title}{Experimental Pulse NMR}}
  (\bibinfo{publisher}{Addison-Wesley Pub. Co}, \bibinfo{year}{1981}).

\bibitem[{\citenamefont{Onak et~al.}(1959)\citenamefont{Onak, Landesman,
  Williams, and Shapiro}}]{onak}
\bibinfo{author}{\bibfnamefont{T.~P.} \bibnamefont{Onak}},
  \bibinfo{author}{\bibfnamefont{H.}~\bibnamefont{Landesman}},
  \bibinfo{author}{\bibfnamefont{R.~E.} \bibnamefont{Williams}},
  \bibnamefont{and} \bibinfo{author}{\bibfnamefont{I.}~\bibnamefont{Shapiro}},
  \bibinfo{journal}{J.\ Phys.\ Chem.} \textbf{\bibinfo{volume}{63}},
  \bibinfo{pages}{1533} (\bibinfo{year}{1959}).

\bibitem[{\citenamefont{Andrew and Tunstall}(1961)}]{andrew}
\bibinfo{author}{\bibfnamefont{E.~R.} \bibnamefont{Andrew}} \bibnamefont{and}
  \bibinfo{author}{\bibfnamefont{D.~P.} \bibnamefont{Tunstall}},
  \bibinfo{journal}{Proc.\ Phys.\ Soc.\ London} \textbf{\bibinfo{volume}{78}},
  \bibinfo{pages}{1} (\bibinfo{year}{1961}).

\bibitem[{\citenamefont{Pavarini and Mazin}(2001)}]{pavarini}
\bibinfo{author}{\bibfnamefont{E.}~\bibnamefont{Pavarini}} \bibnamefont{and}
  \bibinfo{author}{\bibfnamefont{I.~I.} \bibnamefont{Mazin}},
  \bibinfo{journal}{Phys.\ Rev.\ B} \textbf{\bibinfo{volume}{64}},
  \bibinfo{pages}{140504(R)} (\bibinfo{year}{2001}).

\bibitem[{\citenamefont{Belashchenko et~al.}(2001)\citenamefont{Belashchenko,
  Antropov, and Rashkeev}}]{prbant}
\bibinfo{author}{\bibfnamefont{K.~D.} \bibnamefont{Belashchenko}},
  \bibinfo{author}{\bibfnamefont{V.~P.} \bibnamefont{Antropov}},
  \bibnamefont{and} \bibinfo{author}{\bibfnamefont{S.~N.}
  \bibnamefont{Rashkeev}}, \bibinfo{journal}{Phys.\ Rev.\ B}
  \textbf{\bibinfo{volume}{64}}, \bibinfo{pages}{132506}
  (\bibinfo{year}{2001}).

\bibitem[{\citenamefont{Andersen et~al.}(1986)\citenamefont{Andersen,
  Pawlowska, and Jepsen}}]{lmto}
\bibinfo{author}{\bibfnamefont{O.~K.} \bibnamefont{Andersen}},
  \bibinfo{author}{\bibfnamefont{Z.}~\bibnamefont{Pawlowska}},
  \bibnamefont{and} \bibinfo{author}{\bibfnamefont{O.}~\bibnamefont{Jepsen}},
  \bibinfo{journal}{Phys.\ Rev.\ B} \textbf{\bibinfo{volume}{34}},
  \bibinfo{pages}{R5253} (\bibinfo{year}{1986}).

\bibitem[{\citenamefont{Hebel and Slichter}(1959)}]{hebel}
\bibinfo{author}{\bibfnamefont{L.~C.} \bibnamefont{Hebel}} \bibnamefont{and}
  \bibinfo{author}{\bibfnamefont{C.~P.} \bibnamefont{Slichter}},
  \bibinfo{journal}{Phys.\ Rev.} \textbf{\bibinfo{volume}{113}},
  \bibinfo{pages}{1504} (\bibinfo{year}{1959}).

\bibitem[{\citenamefont{Hebel}(1959)}]{hebel2}
\bibinfo{author}{\bibfnamefont{L.~C.} \bibnamefont{Hebel}},
  \bibinfo{journal}{Phys.\ Rev.} \textbf{\bibinfo{volume}{116}},
  \bibinfo{pages}{79} (\bibinfo{year}{1959}).

\bibitem[{\citenamefont{Bud'ko et~al.}(2001{\natexlab{b}})\citenamefont{Bud'ko,
  Kogan, and Canfield}}]{budko}
\bibinfo{author}{\bibfnamefont{S.~L.} \bibnamefont{Bud'ko}},
  \bibinfo{author}{\bibfnamefont{V.~G.} \bibnamefont{Kogan}}, \bibnamefont{and}
  \bibinfo{author}{\bibfnamefont{P.~C.} \bibnamefont{Canfield}},
  \bibinfo{journal}{Phys.\ Rev.\ B} \textbf{\bibinfo{volume}{64}},
  \bibinfo{pages}{180506(R)} (\bibinfo{year}{2001}{\natexlab{b}}).

\bibitem[{\citenamefont{Simon et~al.}(2001)}]{simon}
\bibinfo{author}{\bibfnamefont{F.}~\bibnamefont{Simon}} \bibnamefont{et~al.},
  \bibinfo{journal}{Phys.\ Rev.\ Lett} \textbf{\bibinfo{volume}{87}},
  \bibinfo{pages}{047002} (\bibinfo{year}{2001}).

\bibitem[{\citenamefont{Lee et~al.}(2001)\citenamefont{Lee, Mori, Masui,
  Eltsev, Yamamoto, and Tajima}}]{lee}
\bibinfo{author}{\bibfnamefont{S.}~\bibnamefont{Lee}},
  \bibinfo{author}{\bibfnamefont{H.}~\bibnamefont{Mori}},
  \bibinfo{author}{\bibfnamefont{T.}~\bibnamefont{Masui}},
  \bibinfo{author}{\bibfnamefont{Y.}~\bibnamefont{Eltsev}},
  \bibinfo{author}{\bibfnamefont{A.}~\bibnamefont{Yamamoto}}, \bibnamefont{and}
  \bibinfo{author}{\bibfnamefont{S.}~\bibnamefont{Tajima}},
  \bibinfo{journal}{J.\ Phys.\ Soc.\ Japan} \textbf{\bibinfo{volume}{70}},
  \bibinfo{pages}{2255} (\bibinfo{year}{2001}).

\bibitem[{\citenamefont{Kitaoka}()}]{kitaoka}
\bibinfo{author}{\bibfnamefont{Y.}~\bibnamefont{Kitaoka}},
  \emph{\bibinfo{title}{Private communication}}.

\end{thebibliography}

\end{document}